\documentstyle[twoside,fleqn,espcrc2]{article}
\newcommand{\AmS}{{\protect\the\textfont2
  A\kern-.1667em\lower.5ex\hbox{M}\kern-.125emS}}

\title{Black holes and D-branes
}

\author{Juan M. Maldacena\address{Department of Physics and Astronomy, 
        Rutgers University, \\ 
        Piscataway, NJ, 08855, USA}%
        \thanks{Lectures presented at the Asian Pacific Center for 
Theoretical Physics, December 1996 and at the 33rd Karpacz Winter  
        School of Theoretical Physics, Karpacz, Poland, 13-22 February
1997 
              }
       }

\def\be{\begin{equation}}
\def\ee{\end{equation}}
\newcommand{\la}[1]{\label{#1}}
\def\ba{\begin{array}}
\def\ea{\end{array}}
\newcommand{\refnew}[1]{(\ref{#1})}

\def\sq2{\sqrt{2}}

\def\p{\partial}

\def\RN{Reissner-Nordstr{\o}m}

\def\s42{ 2^{-{1\over 4} } }

\def\g{\gamma}

\def\a{\alpha}
\def\sa{r_0^2 {\rm sinh}^2\alpha }
\def\sg{r_0^2 {\rm sinh}^2\gamma }
\def\ss{r_0^2 {\rm sinh}^2\sigma }

\def\[{\left [}
\def\]{\right ]}
\def\({\left (}
\def\){\right )}

\input epsf.tex

\begin{document}
\bibliographystyle{unsrt}

\begin{abstract}

Black holes have an entropy proportional to the area of the horizon.
The microscopic degrees of freedom that give rise to the entropy are
not visible in the classical theory. 
String theory, a quantum theory of gravity, provides a microscopic
quantum description of the thermodynamic properties of some
extremal and near extremal charged black holes.
The description uses properties of some string theory solitons called 
D-branes. 
In this description, Hawking radiation emerges as a simple perturbative
process. The low energy dynamics of particle absorption and emission
agrees in detail with the semiclassical analysis in the thermodynamic
limit.

\end{abstract}

\maketitle

\section{Introduction}

Under a wide variety of conditions General Relativity predicts
that singularities will develop \cite{hawkingellis}. 
The cosmic censorship hypothesis states that 
under generic physical situations leading to gravitational collapse
the resulting singularities will be covered by an event horizon 
\cite{censor}. This conjecture has not been yet  proved but there
exists great evidence that it is correct \cite{waldaps}.
The area of the horizon is an interesting quantity since it
always increases upon classical evolution \cite{areaincrease}, 
this looks very similar to the second law of thermodynamics.
The analogy became more precise when Hawking showed \cite{hawrad} that
quantum mechanics implies that black holes emit thermal radiation
with a temperature obeying the first law of thermodynamics 
$d M = T_H d S $,  where the entropy is $S = { A_H \over 4 G_N \hbar }$ 
\cite{hawkingentropy}, 
 $M$ is the black hole mass and $A_H$ is the horizon area 
(from now on we set $\hbar =1 $ but keep $G_N \not = 1$).
The area increase law becomes the second law of thermodynamics.
If one includes Hawking radiation, the black hole mass decreases and
so does the area of the horizon, but the total entropy, defined as 
$S = {A_H /4G_N } + S_{rad}$,  increases.
It has always been a puzzle what the degrees of freedom that give rise
to this entropy are. It seems  clear that some quantum gravity will be
necessary
to describe the microstates. 
String theory \cite{REV} is a theory of quantum gravity so one would naturally 
expect that it should give an answer to this question. 
But string theory is defined perturbatively and black holes
involve strong interactions due to their  large mass. Only when
some non-perturbative tools became available \cite{polchinski,daipol}
could 
 precise calculations  be 
 made \cite{sv}. There are, however,  rough counting arguments
that produce the right scaling for the entropy using just string
perturbation theory \cite{lenyspeculations,senstrings,ghjp}.
Our focus will be to explain the calculations that produce precise
results, which are valid presently only for a  subset of all possible
black hole configurations. For other reviews on this subject 
see \cite{horrev}.

We are going to be treating charged black holes. The cosmic censorship
hypothesis gives a bound for the mass of a black hole in terms of
its charge $M \ge Q$ (in appropriate units). The black hole with $M =Q$ is
called
extremal. We will study extremal and near extremal ($M - Q \ll Q$)
black
holes and we focus on black holes in five spacetime dimensions.
In four spacetime dimensions the discussion is similar.
In section 2 
we write  down the classical supergravity solutions we are going to 
describe. In section 3 we present the D-brane description 
of extremal black holes in five dimensions.
In section 4 we study  near extremal black holes, their entropy and their
decay rates and we compare them to the semiclassical results.
In section 5 we compute the greybody factors for emission of
massless scalars.

\section{Classical solutions}

We  consider type IIB string theory compactified on $T^5$.
The low energy theory is the maximally supersymmetric supergravity
theory in five dimensions. It has 32 supersymmetry generators and
it is the dimensional reduction of ten dimensional type IIB
supergravity. 
The theory contains 27 abelian gauge fields. The full string 
theory contains charged objects  that couple to each of these 
gauge fields. These objects are: 5 Kaluza Klein momenta, 5 string
winding directions, 5 D-string winding directions, 10 possible
D-3-brane wrapping modes, a solitonic NS fivebrane, a D-fivebrane.
All these charges are interchanged by U-duality transformations
\cite{hull}
and they
are all quantized.
Therefore,
 we measure charges in integer   multiples of the elementary units.

We consider a black hole solution which has $Q_5$ D-fivebranes
wrapped on $T^5$, $Q_1$ D-1-branes wrapped on an $S_1$ (we choose
it as the direction $\hat 9$), and momentum  $P = N/R$ also 
along the direction of the D-string (direction $\hat 9$).
When we mention D-branes in the context of classical
solutions
we only refer to the charge that the solution is carrying, there
are no explicit D-branes in the sense of \cite{daipol,polchinski}
anywhere in spacetime. 
We choose this set of charges because the string theory description
is simpler. Other black holes are related to this by U-duality
transformations. Further details about supergravity solutions
can be found in R. Khuri's contribution to this volume
\cite{khuripoland}.

We start by presenting the ten dimensional solution \cite{hms}
\begin{equation} 
\label{dil} 
 e^{-2( \phi -\phi_\infty) } = 
 \(1+   { \sg \over r^2 } \) \(1 + {\sa\over r^2 } \)^{-1}~,
\ee
\begin{equation} 
\la{metric}\ba{l}
ds^2_{str} = 
 \( 1 + { \sa \over r^2}\)^{-1/2} \( 1 + { \sg \over r^2}\)^{-1/2}\times \\
\left[ - dt^2 +dx_9^2 +
{r^2_0  \over r^2} (\cosh \sigma dt + \sinh\sigma dx_9)^2 \right. \\
 +\left. \( 1 + {\sa \over r^2}\) (dx_5^2 + \dots + dx^2_8) \] \\
 + \( 1 + { \sa \over r^2}\)^{1/2}\( 1 + { \sg \over r^2}\)^{1/2}
\times 
\\
 \left[
\(1-{r_0^2 \over r^2}\)^{-1} dr^2 + r^2 d \Omega_3^2 \right]~.
\ea \ee
Also some components of the Ramond-Ramond three-form field strength
$H'_{\mu\nu\rho}$
are nonzero since the solution carries D1-brane and D5-brane charge.
This solution is parameterized by the four  independent quantities
$\alpha,\g,\sigma,r_0$. There are two extra parameters 
which enter through the charge quantization conditions which are
 the radius of the 9$^{th}$ 
 dimension $R_9$ and the product
of the radii in the other four compact directions 
$V \equiv  R_5 R_6 R_7
R_8 $. The three charges are
\be\la{charges}
\ba{rl}
   Q_1 &= {V\over 4\pi^2 g}\int e^{2 \phi_6} *H'
   = { V r_0^2  \over 2 g } \sinh 2 \a , \\
   Q_5 &= {1\over 4\pi^2 g} \int H'  =  { r_0^2\over 2g} \sinh 2 \g ,
\\
  N &= {  R^2V r_0^2 \over 2  g^2} \sinh 2 \sigma ,
\ea \ee
where $*$ is the Hodge dual in the six dimensions $x^0,..,x^5$.
For simplicity we  set from now on $\alpha'=1$. 
All charges are normalized to be integers.
We have chosen a convention such that $g \to 1/g$ under S-duality.
Further explanations on the charge quantization conditions can be
found
in \cite{polchinski,jmmthesis}.

Reducing \refnew{metric}  to five dimensions using the standard 
dimensional reduction procedure \cite{dimensionalreduction},
 the solution takes
the  simple and symmetric form:
\be \la{solnfd} ds_5^2 =  - \lambda^{-2/3}h dt^2 + 
\lambda^{1/3}
\({  dr^2\over h} + r^2 d \Omega_3^2 \right)~,
\ee
where
\be \la{deff} \ba{rl}
 \lambda  
= &\(1+{r_1^2\over r^2} \)\(1+{r_5^2\over r^2} \)\(1+{r_n^2\over r^2} \)~,
\\
h =& 1 - {r_0^2 \over r^2}
\ea
\ee
\be\la{anglesdef}
r_1^2 = \sa~,~~~~r_5^2 = \sg ~,~~~~r_n^2 = \ss 
\ee

This is just the five-dimensional Schwarzschild metric with the time
and space components rescaled by different powers of $\lambda$. 
The event horizon is  at $r=r_0$.
Several thermodynamic quantities can be associated to
this solution. They can be computed in either the ten dimensional or
five dimensional metrics and yield the same answer. For example, the
ADM energy is 
\be \la{mss}
M=  {   R V r_0^2  \over 2  g^2}
(\cosh 2 \alpha + \cosh 2 \gamma + \cosh 2 \sigma  )~.
\ee

The Bekenstein-Hawking entropy is
\be \la{entropy} \ba{rl}
S = &{A_{10}\over 4 G^{10}_N} = {A_5\over 4 G^5_N} \\
 =& { 2 \pi  R V  r_0^3 \over  g^2 } 
 \cosh \alpha \cosh \gamma \cosh \sigma.
\ea \ee
where $A$ is the area of the horizon and
we have used that the Newton constant is $G_N^{10} = 8 \pi^6 g^2 $.
The Hawking temperature is
\be \la{thwk}
T= {1 \over 2 \pi  r_0 \cosh\alpha \cosh \gamma \cosh \sigma } .
\ee
The extremal limit 
corresponds to $r_0 \to 0$, $ \alpha, \gamma,
\sigma
\to \infty$ keeping the charges  \refnew{charges} finite.
In that limit the entropy \refnew{entropy}  becomes 
\be \la{entroext}
S = 2 \pi \sqrt{ N Q_1 Q_5 }
\ee
and the temperature vanishes. 
Note that the extremal entropy is independent
of
any continuous parameters \cite{topological,larsen}.
The extremal black hole backgrounds preserve some space time
supersymmetries and therefore they are  BPS states.
In this case the cosmic censorship bound 
becomes identical to the supersymmetry BPS bound \cite{kalloshpeet}.

 The near extremal limit  corresponds to  $r_0$ 
small
and $\alpha, \gamma, \sigma$  large. 
The relative values of $\alpha, \gamma , \sigma $ are related to 
the total contribution of the different charges to the mass \refnew{mss}.
The near extremal black holes
  that are easiest to  analyze  in terms
of D-branes are  those where  $\sigma \ll \alpha, \gamma$, or
$r_0,r_n \ll r_1, r_5$, which
means that the contribution to the mass \refnew{mss} due to the 
D-branes is much bigger than the contribution due to the 
momentum excitations. This limit is called ``dilute gas'' \cite{hstro}
\cite{asjm}.
In this limit, the mass and entropy  of the near extremal black hole
become
\be \la{dilutegasmass}
 M =  {Q_5 R V \over g} + {Q_1 R \over g } + { R V r_0^2 
 \over 2 g^2 }
 \cosh 2 \sigma ~,
\ee
\be \la{dilutegasentropy}
 S =  2 \pi {R \sqrt{V} r_0 \over g} \sqrt{Q_1 Q_5}  \cosh \sigma ~.
\ee
Note that the five dimensional  \RN\ solution corresponds to the case of
$\alpha = \gamma = \sigma $  which is not included in the
dilute gas limit.

All these black hole solutions will be well defined if curvatures are
everywhere much smaller than $\alpha'$, since otherwise 
 $\alpha' $ corrections to the low energy action become important.
This generically implies that the sizes of the black hole should obey 
$r_1,r_5
,r_n \gg 1 $ (remember that $\alpha'=1$).  The precise condition
will be a little more complicated if some scale is very different 
from the others.
If the compactification sizes are of the order of $\alpha' $ this
implies that
\be\la{bigq}{
gQ_1 \gg 1,~~~~g Q_5 \gg 1,~~~~g^2 N \gg 1 
}
\ee
Note that we cannot enforce \refnew{bigq} by seting $Q=1$ and making
$g \gg 1 $ since the condition \refnew{bigq} was derived in 
weakly coupled string theory, if $g \gg 1 $ the corresponding bound
comes  from light D-strings and implies again that $Q \gg 1$.
Therefore  black holes always involve large values of the 
charges.

\section{D-brane description of extremal 5d black holes}

We continue with  type IIB string theory on $T^5 = T^4\times S^1$.
We consider a configuration of
$Q_5$ D-fivebranes wrapping the whole $T^5$,
$Q_1$ D-strings wrapping the $S^1$ and momentum $N/R_9$ 
along the $S^1$, choosing this $S^1$ to be in the
direction $\hat9$. All charges $N,Q_1,Q_5$ are integers.
For a review of D-branes see \cite{polchinskinotes}.
We take the coupling constant $g$ to be small and the radius
$R_9$ to be large. 
The total mass of the system is 
\be \la{bpsmass}
M = { Q_5 R V \over g} + { Q_1 R \over g } + { N \over R }
\ee
and it saturates the corresponding BPS bound.

We will calculate the entropy of this state in perturbative string 
theory. This calculation was first done by Strominger and Vafa
\cite{sv}.
Since extremal D-branes are boost invariant along the
directions parallel to the branes they cannot carry
momentum along $S_1$ by just moving rigidly.
Our first task will be to identify the D-brane excitations
that carry the momentum. 
The BPS mass formula for the whole system implies that
these excitations have to be massless and moving along the
$S^1$ since the excitation energy, defined as 
the total mass of the system  minus the 
mass of the onebranes  and fivebranes, is equal to the momentum.
 If any excitation fails to be massless it
would contribute more to the energy than to the momentum and
the BPS mass formula would be violated. 
Excitations of the 
branes are described by massless open strings.
There are many types of open strings to consider: those that
go from one 1-brane to another 1-brane, which  we denote  as (1,1) strings,
as well as the corresponding  (5,5), (1,5) and (5,1) strings (the last two
being different because the strings are oriented). 
We want to excite these strings and make them carry momentum
in the direction of the $S^1$. 
Exciting some of them makes others massive \cite{jmmthesis} so we
have to find  the way to excite the stings so that
a maximum number remains massless, since  this  configuration will
 have the 
highest entropy. 
Let us start by  working
 out the properties of (1,5) and (5,1) strings.
The string is described by the usual string action where 
two of the coordinates have Neumann-Neumann boundary conditions 
($X^0,X^9$), the four extended spatial 
coordinates have Dirichlet-Dirichlet boundary
conditions ($X^1,X^2,X^3,X^4$) and the other four internal coordinates
have
Neumann-Dirichlet conditions ($X^5,X^6,X^7,X^8$).
The vacuum energy of the worldsheet bosons is 
$ E = 4(-1/24 + 1/48)$. Consider the NS sector for the
worldsheet fermions, 
the  4 that
are in the ND directions will end up having R-type quantization
conditions. The net fermionic vacuum energy is $ E= 4( 1/24 -  1/48) $
and exactly cancels the bosonic one.
This vacuum is a spinor under SO(4)$_{5678}$
 and obeys the GSO chirality condition
$ \Gamma^5\Gamma^6\Gamma^7\Gamma^8 \chi = \chi$. What remains is a
two dimensional representation. There are
two possible orientations and they can be attached to
any of the different branes of each type. This gives a total
of $ 4 Q_1 Q_5$ different possible states for these strings.
Now consider the Ramond sector, the four internal fermions
transverse to the string will have NS type boundary conditions. The
vacuum again has zero energy and is an SO(1,5)$_{091234}$ spinor and, 
therefore, a  
spacetime  fermion.
Again the GSO condition implies that only the positive
chirality representation of SO(1,5)$_{091234}$ survives.
When it is also  left moving  only the $ 2^+_+$ under
SO(1,1)$_{09}\times$SO(4)$_{1234}$ survives. 
This gives the same number of states as
for the bosons.
Note that the fermionic (1,5) (or (5,1)) strings carry angular
momentum under the spatial rotation group SO(4)$_{1234}$ but the 
bosonic (1,5) (or (5,1)) strings do not carry angular momentum.

\vskip 1cm
\vbox{
{\centerline{\epsfxsize=3in \epsfbox{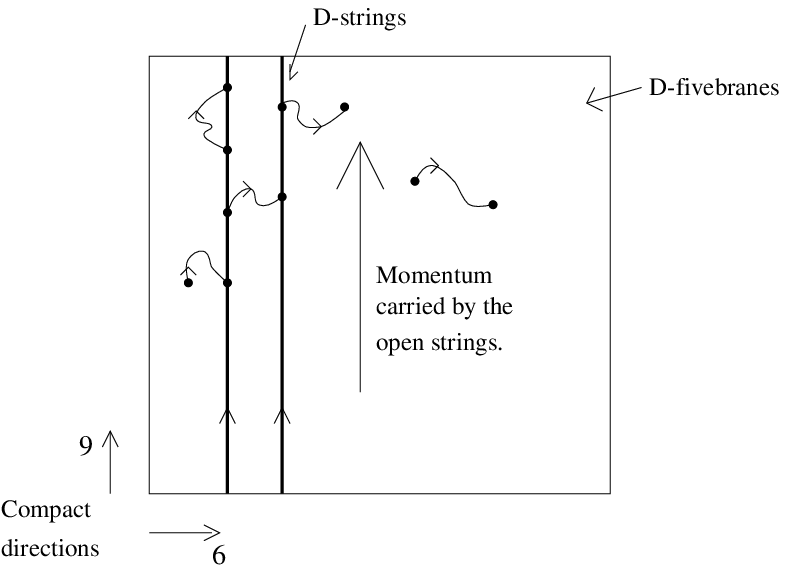}}}
{\centerline{ FIGURE 1:
Configuration of intersecting D-branes.}}
{\centerline{Open strings go between different branes}}
}
\vskip .5cm

The (1,1), (5,5) and (1,5) strings interact among themselves.
We are interested in the low energy limit of these interactions.
This corresponds to the field theory limit of the system of
branes. If we take the size of the S$_1$ along $\hat 9$  to be 
very large this will be a 1+1 dimensional gauge theory.
The Lagrangian is then determined by supersymmetry
and gauge invariance. The (1,1) and (5,5) strings are
U($Q_1$) and U($Q_5$) gauge bosons respectively and the
(1,5) strings (and (5,1) strings) are in the fundamental
(antifundamental) 
of U($Q_1$)$\times$U($Q_5$). 
This Lagrangian was studied by Douglas \cite{dgl} and it turns out
that after these interactions are taken into account only
$4Q_1Q_5$ truly  massless degrees of freedom remain.
In gauge theory terms, one is interested in the Higgs branch
of the theory which is $4Q_1Q_5$-dimensional. 
We will spare the details which can be found in \cite{jmmthesis} ~(page 52),
and  conclude that the number of massless states is 
$4Q_1Q_5 $ (with the same number of bosonic and fermionic 
states since the theory on the branes is supersymmetric). These 
massless degrees of freedom are described by a two dimensional 
conformal field theory \cite{sv}, which accounts for the 
low energy excitations of the D-brane system. In fact, it is a (4,4) 
superconformal field theory, i.e. it has four left moving and four
right moving supersymmetry generators. The rotational symmetry
SO(4)$_{1234}
 \sim \ $
SU(2)$_L\times$ SU(2)$_R$ of the four spatial dimensions 
acts on this superconformal field theory as the 
SU(2)$_L\times $SU(2)$_R$ R-symmetries of the N=4 supersymmetry
algebra in two dimensions. Notice that the chirality in space becomes
correlated with the chirality of the 1+1 dimensional theory. 
So that SU(2)$_L$ in spacetime acts on the leftmovers and similarly
for SU(2)$_R$ \cite{spn} .

The BPS state that we are interested in has only left moving 
excitations so the rightmovers are in their ground state $E_R=0$,
The state 
counting is the same as that of 
 a one dimensional gas of left moving particles  with
$ N_{B,F} = 4 Q_1 Q_5 $
 bosonic and fermionic species with  energy 
$E=E_L=N/R_9$
on a compact one dimensional space of length $L= 2 \pi R_9 $.
The standard entropy formula gives \cite{sv} \cite{cama}
\be\la{entropyd}
 S_e = \sqrt{ \pi (2  N_B + N_F) E L/6 }   
=2 \pi\sqrt{ Q_1Q_5 N}~, 
\ee
in perfect agreement with \refnew{entroext},
 including the numerical coefficient.
It might seem  surprising that a system could have entropy at
zero temperature, but this is a common phenomenon.
Consider for example a gas of massless particles in a box with
periodic
boundary conditions constrained to 
have a fixed amount of momentum, at $T=0$ the entropy remains
nonzero. 
This reason is  {\it exactly} the same reason that 
 black hole 
entropy is nonzero at $T=0$.

In our previous argument we implicitly  took
the D-strings and the fivebranes to be singly wound since we were
 assuming that the excitations  carried momentum quantized in units of
$1/R$.
For large $N$, $N \gg Q_1 Q_5 $,  the entropy \refnew{entropyd} is the
same no matter how the branes are wound. 
However for $ N \sim Q_1 Q_5  $ the winding starts to matter.
The reason is that in order for the asymptotic entropy
formula to be correct for low $N$ we need to have
enough states with small energies \cite{sm}. 
Let us study the effect of different wrappings.
We first simplify the problem and 
 consider a set of $Q_1$  1-branes wrapped on $S^1$, ignoring for
the
time being, the 5-branes. We may distinguish the
various ways
the branes interconnect. For example,
 they may connect up so as to form one
long
brane of total length $R'= R Q_1$. At the opposite extreme they might
form $Q_1$ disconnected loops. The spectra of open strings is different in
each
case. For the latter case the open strings behave like $Q_1$ species of
1 dimensional particles, each  with energy spectrum given by integer
multiples of
$1/R$. In the former case they behave more like a single species of 1
dimensional
particle living on a space of length $Q_1 R$. The result
\cite{dasmathur}
 is a
 spectrum
of
single
particle energies given by integer multiples of $1\over{Q_1 R}$  .
 In other
words
the system simulates a spectrum of   fractional charges. For consistency the
total
charge must add up to an integer multiple of $1/R$ but it can do so by
adding
up
fractional charges.

Now let us return to the case of both 1 and 5 branes. By suppressing
reference
to
the four compact directions orthogonal to $x^9$ we may think of
the 5 branes
as
another kind of 1 brane wrapped on $S^1$. The 5-branes
 may also be connected
to
form a single multiply wound brane or several singly wound branes. Let us
consider the spectrum
 of (1,5) type strings (strings which connect a 1-brane
to
a five-brane) when both the 1 and 5 branes each
form a single long brane. The 1-brane has total length $Q_1 R$ and the
5-brane
has length $Q_5 R$. A given open string can be indexed by a pair of indices
$[i,\bar j]$ 
labeling which loop of 1-brane and 5-brane it ends on. As a
simple
example choose $Q_1=2$ and $Q_5=3$. Now start with the $[1,1]$ string which
connects the first loop of 1-brane to the first loop  of 5-brane. Let us
transport this string around the $S^1$. When it comes back to the starting
point
it is a $[2,2]$ string. Transport it again and it becomes a $[1,3]$
 string.
It
must be cycled 6 times before returning to the $[1,1]$ configuration. It
follows
that such a string has a spectrum of a single species living on a circle of
size
$6R$. More generally, if $Q_1$ and $Q_5$ are relatively prime the system
simulates a single species
 on a circle of size $Q_1 Q_5 R$. If the $Q's$ are
not
relatively prime the situation is slightly
more complicated but the result is the same. 
A more detailed picture of how this happens is presented
in \cite{wadia}.


We can easily see that this way of wrapping 
the branes  gives the correct value for the extremal
entropy. 
 As above,   the open strings have 4 bosonic
and 4 fermionic degrees of freedom and carry total
momentum $N/R$. This time the quantization length is
$R'=Q_1Q_5R$ and the momentum is quantized in
units of $(Q_1 Q_5 R)^{-1}$. Thus instead of being at level
$N$ the system is at level $N'=NQ_1Q_5$. In place of the original
$Q_1Q_5$
species we now have a single species. 
The result is
\be\la{entrocorr}
S=2\pi\sqrt{N'}=2\pi\sqrt{NQ_1Q_5}
\ee
So we have a long effective string that is moving along
the fivebrane. In the extremal case, this effective string
picture follows precisely from an analysis of the moduli
spaces of BPS states \cite{mooreverlinde}.
What one actually has is a sum over multiple string states
 which one can call ``second quantized'' 
strings on a fivebrane \cite{verlinde}.
 The state in which they are all connected
into a single long string is the one having most  entropy.

For completeness we will now present the same calculation but in
a picture where we start with just D-fivebranes and we build up
the charges as excitations of the D-fivebranes. 
We start with $Q_5$ D-fivebranes. The low energy theory on the
fivebranes is  an  U($Q_5$) supersymmetric Yang Mills with 16 
supersymmetries (same amount as N=4 in d=4). 
This theory contains BPS string solitons with are constructed as 
follows: if the fivebranes are along the directions 56789, 
take an instanton configuration that involves the 
directions  5678 and the gauge fields along those directions.
The corresponding field configuration could be localized in the
directions 5678 but will be extended along the direction 9, so it
is a string soliton.
Notice that, even though we call this solution an ``instanton'' in
the sense that the Yang-Mills fields are self dual 
solutions of a YM theory in four dimensions (5678), the physical
interpretation is that we have a string ``soliton'' which exists for
all times. 
It turns out that each instanton that lives on the fivebrane world
volume carries one unit of D-string charge \cite{dgl} due to a 
Chern Simon coupling on the fivebrane of the 
form 
\be\la{coupling} 
 \ ~~~~ \int d^{1+5 }x \  B_2^{RR} \wedge F \wedge F 
\ee
since $F\wedge F$ will be proportional to the instanton number.
We are interested in the case that  the instanton number is $Q_1$.
So D-strings disolve into instantons when they get into fivebranes.
In fact, giving an expectation value to the (1,5) strings corresponds
to giving a size to the instanton \cite{dgl,wittensmall}.
This description makes sense, in principle, only  when the mass of the
fivebranes is much bigger than the mass of the D-1-branes in
\refnew{bpsmass},
since otherwise the fivebranes  would contain so much energy  that
they  would   no longer be   described by the low energy YM theory. 
This instanton configuration
is characterized by 4$ Q_1Q_5$ continous parameters
which specify the intanton  positions on the branes as well as their 
relative orientation inside the U($Q_5$) gauge group, the  space
of these  parameters is called ``moduli space''. 
When we put some momentum along the direction $\hat 9$, this momentum can
be
carried by small oscillations of the instanton configuration.
We denote the  instanton parameters  by $\xi^a$, $a = 1, .., 4
Q_1Q_5$.
They can be 
 slowly varying functions $\xi^a(t- x^9 ) $
representing traveling waves moving along the instantons. 

These are  small 
oscillations in the parameters specifying the instanton configuration,
i.e. oscillations in moduli space.
Each bosonic mode has a fermionic superpartner and together they form
a (4,4) superconformal field theory with 
central charge $c = 6 Q_1 Q_5$. 
A configuration carrying momentum $N$ corresponds to states in
the SCFT with $L_0 = N$ and $\bar L_0 =0$, the entropy of such states
can be calculated using the CFT formula $d(N) \sim e^{ 2 \pi \sqrt{ N
c/6}}$
for the degeneracy of states at level $N$. This yields \refnew{entropyd}
again.
The moduli space of instantons is, topologically,
a symmetric product: ${\cal M } = (T^4)^{Q_1Q_5}/S(Q_1Q_5)$ 
\cite{vafacount,vafasadov}.
This is the target space of the SCFT, in other words:
 $\xi^a(t, x^9) $
defines a map from $R \times S_1$ to ${\cal M}$.
Since 
we have twisted sectors there are low lying 
modes with energies of the order of $1/RQ_1Q_5$ which give rise to
the long effective string picture described above.
This picture where we start with only one kind of branes is the one
that we would naturally use \cite{martinec,verlindebh}
if we are working in the M(atrix) theory
of  \cite{bfss}.

We should also address  the question of whether the system really forms
a bound state, this is a question on the behavior of
the zero modes on the moduli space. The analysis of \cite{sv}
shows that they indeed form a bound state.
Note that also for entropic reasons the state would  stay
together
for a long time. 

It is  interesting that when the momentum is not
uniformly distributed along the string (the $\hat 9$ direction) the
D-brane calculation still agrees with the corresponding black hole
result \cite{marolf}.

In performing these calculations we have assumed that the coupling was
weak. One might naively think that the only condition is $g\ll 1$. 
However, since there is a large number of branes there could be 
large $N=Q$ effects which grow as $g Q$. At low energies these are
just
the large $N$ effects of the Yang-Mills theory (note that the YM
coupling
is $g_{YM } = \sqrt{g}$). In string theory they correspond to the 
possibility of inserting a hole on the worldsheet, each hole has a
power
of $g$ and a factor of $Q$ coming from the trace over Chan Paton
factors. So the above calculations are correct when 
\be \la{weakcoupling}
\ ~~~~~~~~~~~~gQ \ll 1
\ee
On the other hand the classical black hole solution is well defined
when $g Q \gg 1 $.
The reason that we expect agreement (and {\it find } it) is
that we are counting the number of BPS bound states and this number is not
expected to change as we vary the coupling constant. Notice that 
indeed the extremal black hole entropy \refnew{entroext} 
is independent of the coupling
and all other continuous parameters.

\section{Near Extremal black holes}

We now turn to a discussion of near  extremal five-dimensional
black holes \refnew{solnfd} \refnew{dilutegasmass}\refnew{dilutegasentropy}.
For the reasons that we have just discussed we might naively  not expect 
agreement in this case. However, in the dilute gas regime, we are very
close to a configuration of extremal D1 and D5 branes and
supersymmetry
non-renormalization arguments do indeed help us \cite{jmmlow,dasloop} and 
explain the agreement that we are going to find. 
We are going to consider a weakly coupled system of D-branes but
we will always restrict to the low energy approximation.
 We will see that for the near-extremal case the
 agreement between the two approaches
is just as impressive as in the extremal case. 
The D-brane model is 
 a low energy approximation to the full quantum dynamics of 
black holes. The energy should be low compared to the scale set by
gravitational size of the black hole $r_s$ defined as the radius 
at which the redshift of a static observer becomes of order one,
$r_s^2  \sim g Q \sim r_1^2, r_5^2 $. 
The condition on the energy becomes $\omega r_s \ll
1$, where $\omega$ represents the typical energy of the
brane excitations, as well as the Hawking temperature of the
system \cite{jmmlow}. We will not go into the details of the 
justification of this extrapolation which can be found in
\cite{jmmlow}.

We start with a system of 1D-branes and 5-D branes as before and
we add some extra energy and momentum to the system.
This energy excites the  massless left and right moving modes of
the instanton configuration.
In the instanton picture we 
say that we are creating left and right moving excitations on the  moduli 
space, $\xi^I(t,x^9)$.
Using the black hole formulas \refnew{charges},\refnew{dilutegasmass}  we
calculate the energies of the left and right movers
\be\la{masslefri}\ba{rl}
M - { Q_5 RV \over g} - {Q_1 R \over g} =&
{ RV r_0^2 \over 2 g} \cosh 2 \sigma \\=& { N_L + N_R \over R }~,
\ea
\ee
\be\la{chargelefri}
N = {R^2 V r_0^2 \over 2g^2 }\sinh 2 \sigma = N_L -N_R ~.
\ee
In this fashion we can calculate $N_{L,R}$ in terms of the black
hole parameters. 
The entropy calculation proceeds as in the extremal case. We 
work in the multiply wound picture with 
 4 bosons and 4
fermions with effective $N_{L,R}' = Q_1Q_5 N_{L,R}$.
We find that 
\be\la{entronear}\ba{rl}
 S= &2\pi(\sqrt{N'_L}+ \sqrt{N'_R}) \\
=&
 2\pi \sqrt{Q_1Q_5 N_L} + 2 \pi \sqrt{Q_1Q_5 N_R }~.
\ea
\ee
We see that this result agrees with the near extremal entropy in 
the dilute gas limit \refnew{dilutegasentropy} once we use \refnew{masslefri}
\refnew{chargelefri}. 
This is the simplest case, if we want to
consider more general near extremal black holes, including
\RN, one  has
to include other excitations besides the right movers \cite{cama,truchos} and
the arguments are not so well justified.

These non-BPS states will decay. The simplest decay process is
a collision of a right moving excitation with a
left moving one to give a closed string mode that leaves the
brane. We will calculate the  emission rate for
 uncharged  particles. The 
basic process is a
 right moving mode  with momentum $p_9 = n/R_9Q_1 Q_5$ colliding
with a left moving one of momentum $p_9 = - n/R_9Q_1 Q_5$ to give a closed
string mode of energy $\omega = 2 n/RQ_1 Q_5 $. 
Notice that we are considering the branes to be multiply wound
since that is the configuration that had the highest entropy. 
If the momenta are not  exactly
opposite the outgoing string  carries some momentum in the
$9^{th}$ direction and  we get  a charged
particle from the five dimensional point of view.
 Notice that the momentum in the $9^{th}$ direction of
the outgoing particle has to be quantized in units of $1/R_9$, only
particles
on the branes can have fractional momenta!.  
This means that outgoing charged particles have a very large mass,
and that they are  thermally suppressed when $R_9$ is small.
 All charged 
particles have  masses
 of at least the compactification scale.
\vskip 1cm
\vbox{
{\centerline{\epsfxsize=3in \epsfbox{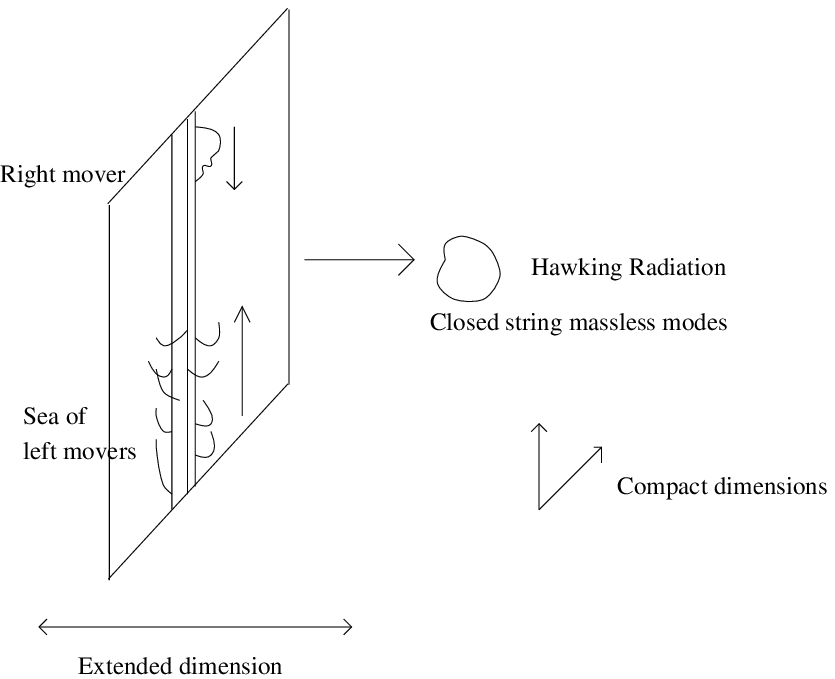}}}
{\centerline{ FIGURE 2:
D-brane picture of the Hawking }}
{\centerline{radiation emission process.
}}
}
\vskip .5cm

In other words, we have a very long effective string winding around
 the
compact direction $\hat 9$, it can oscillate along the other 4 compact
dimensions (5678)
and it  emits gravitational quadrupole radiation. The graviton
is polarized along the compact directions and is  a scalar
from the point of view of the five dimensional observer.

We will calculate the rate for this process according to
the usual rules of relativistic quantum mechanics and show
that the radiation has a thermal spectrum if we do not
know the initial microscopic state of the black hole.

The state of the D-branes  is specified by giving the
left and right moving occupation numbers of each of the
 bosonic and fermionic oscillators.
In fact, the near extremal D-branes  live in
a subsector of the total Hilbert space that is
isomorphic to  the Hilbert space of a 1+1 dimensional
CFT.
The initial  state $|\Psi_i\rangle$ can  emit  a
closed string mode and become $|\Psi_f\rangle$.
The rate, averaged over initial states and summed over final states
(as one would do for calculating the decay rate of an unpolarized
atom)  is
\be\la{rateunpol} \ba{rl}
d \Gamma \sim & { d^4  k \over \omega } { 1 \over p_0^R p_0^L V R }
\delta( \omega -( p_0^R + p_0^L) ) \times 
\\ &\sum_{i,f}
\left| \langle \Psi_f| H_{int} | \Psi_i \rangle  \right|^2
\ea \ee
We have included the factor due to 
the compactified volume  $R V$.  
The relevant string amplitude for this process is given
by a correlation function on the disc with two insertions
on the boundary, corresponding to the two open string
states and an insertion in the interior, corresponding to
the closed string state.
We consider the case when the outgoing closed string is
a spin zero  boson in five dimensions,
so that it corresponds to the dilaton,  the
internal metric,  internal $B_{\mu\nu}$ fields,
or internal components of RR gauge fields.
This disc  amplitude, call it  $\cal A $,  is proportional to the
string coupling constant $g$
and to $\omega^2$ \cite{hk}. The reason for this
last fact is that it has to vanish when we go to zero
momentum, otherwise it would indicate that there is
a mass term for the open strings (since one can vary the vacuum
expectation value of the corresponding closed string fields
continuously). 
In conclusion, up to numerical factors,
\be\la{stringamplitude}
 { \cal A} \sim g \omega^2.
\ee

Note that performing the average over initial and summing
 over final states
will just produce a factor of the form $ \rho_L(n) \rho_R (n)$ with
\be\la{rhoright}
\rho_R (n) = {1 \over N_i} \sum_{i} \langle \Psi_i | a^{R \dagger}_n
 a^R_n | \Psi_i
\rangle
\ee
where $N_i$ is the total number of initial states and
$ a_n^R $ is the creation operator for one of the $ 4  $
bosonic open
string states. The factor $ \rho_L(n) $ is similar.
Since we are  just
averaging
 over all possible initial states with given value of $N_R$,
this corresponds to taking the expectation value of $a^\dagger_n a_n$ in
the microcanonical ensemble with total energy $E_R = N_R/R_9 =
N'_R / R_9 Q_1 Q_5 $ 
of a one
dimensional gas. Because  $ N'_R  $ is large compared to one,
 we can calculate \refnew{rhoright} in
the canonical ensemble.
The occupation number is then
$$ \rho_R(\omega)  = { e^{- \omega  \over 2 T_R } \over
1 - e^{ - \omega \over 2 T_R } }~.
$$
We can read off the ``right moving'' temperature
\be\la{tright}
 T_R = {1 \over \pi}   {1 \over R }
 \sqrt{  N_R \over Q_1 Q_5 }~.
\ee
There is a similar factor for the left movers $\rho_L$
with a similar looking temperature
\be\la{tleft}
 T_L = {1  \over \pi}   {1 \over R }
 \sqrt{  N_L \over Q_1 Q_5 }~.
\ee
In fact it would be more accurate to say that there is only
one physical  temperature  of the gas, which 
agrees with the Hawking temperature of the corresponding black hole,
\be\la{thawk} 
{ 1\over T_H } = {1 \over 2} ( {1 \over T_L } +
{1 \over T_R } ) 
\ee
and that $T_{L,R}^{-1} = T_H^{-1} ( 1 \pm \mu )$
 are some natural combination of the temperature and
the chemical potential, which 
gives the gas some net momentum. 
Using the values for $N_{L,R}$ from \refnew{masslefri} \refnew{chargelefri}  
we find 
\be\la{tlbhparam}
T_L = {1\over \pi}{r_0 e^\sigma \over 2 r_1 r_5 } ~,~~~~~~~~~~~~~~~~~
T_R = {1\over \pi}{r_0 e^{-\sigma} \over 2 r_1 r_5 }~.
\ee
The expression for the rate  is, up to a numerical constant,
\be\la{ratenew}
d\Gamma   \sim  { d^4 k \over \omega  } { 1 \over p_0^R p_0^L R V }
|{ \cal A}|^2  Q_1 Q_5 R \rho_R(\omega)  \rho_L(\omega)
\ee
where ${\cal A}$ is the disc diagram result.
 The factor
 $Q_1 Q_5 R$ is a volume factor, which 
arises from the delta function of momenta in \refnew{rateunpol}
$ \sum_n \delta (\omega  - 2 n/RQ_1Q_5 ) \sim R Q_1 Q_5 $.
The final expression for the rate is, using
\refnew{stringamplitude} in \refnew{ratenew}, 
\be\la{ratefinal}
 d\Gamma =  {  \pi^3 g^2 \over  V } Q_1Q_5 \omega  
{ 1 \over e^{\omega \over 2 T_L} -1 }{ 1 \over e^{\omega \over 2 T_R} -1 }
{d^4 k \over (2\pi)^4}
\ee
We have not  shown here how to calculate the precise numerical
constant in front of \refnew{ratefinal} , this precise calculation
was done in  \cite{dasmathurone}, and we refer the
reader to it for the details. 

If we are considering a black hole which is very close to
extremality  with nonzero momentum $N \sim N_L \gg N_R$
then we find  from \refnew{tleft}\refnew{tright} that $T_L \gg T_R$.
Examining the expression for the rate \refnew{ratefinal} we see that the
typical emitted energies are  of the order of $T_R$.  Therefore,
we can approximate the left moving thermal factor by
\be\la{rholeft}
\rho_L \sim { 2T_L\over \omega }
\ee
 and replacing it in \refnew{ratefinal} we 
find
\be\la{ratearea}\ba{rl}
 d\Gamma =& { 2 \pi^2 g^2 \over R V } \sqrt{Q_1Q_5 N} 
{ 1 \over e^{\omega\over 2 T_R} -1 }{d^4 k \over (2\pi)^4} \\ 
= &
A_H { 1 \over e^{\omega \over 2 T_R} -1 }
{ d^4 k \over (2\pi)^4}
\ea \ee
where $A_H$ is the area of the horizon. 
We conclude that the emission is thermal, with a physical
Hawking temperature
\be\la{hawkingstring}
 T_H = 2 T_R 
\ee
which exactly matches the classical result \refnew{thwk}.
The area appeared correctly in
\refnew{ratefinal} \cite{wadiaarea} . 
Actually,   the coupling constant coming from
the string amplitude $\cal A$   is hidden in the
expression for the area (area = $4 G_N^5 S $). The overall
coefficient in \refnew{ratearea} matches precisely with the semiclassical
result \cite{dasmathurone}.

Notice that if we were emitting a spacetime fermion then
the left moving mode  could be a boson and the right moving
mode  a fermion, this produces the correct thermal factor
for a spacetime fermion. The opposite possibility gives
a much lower rate, since we do not have the enhancement
due to the large $\rho_L$ \refnew{rholeft}.

When separation from extremality is very small, then the number
of right movers is small and the statistical arguments used
to derive \refnew{ratefinal} fail. Semiclassically this should happen when
the temperature is so low that the emission of one quantum at temperature
$T$ causes the temperature to change by an amount of order $T$. This
means that the specific heat is of order one. This happens when
the 
 mass difference from extremality is \cite{limitations}  
\be\la{gapclass}
\delta M_{min} = M-M_{BPS} \sim { {G_N^5} \over  r_e^4 }
\ee
for a \RN\ black hole, with  $r_e$ being the Schwarzschild radius
of the solution. 
The D-brane approach suggests the
existence of a mass gap 
\be\la{gapdbrane}
\delta M_{min} \sim  
{2\over  Q_1 Q_5 R  }
\ee
which using  \refnew{charges}  scales like \refnew{gapclass}.
This is an extremely  
 small energy for a macroscopic extremal black hole. In
fact,
 it is of the order of the kinetic energy that the black hole would
have, due to the uncertainty principle, if we want to measure its
position with an accuracy of the order of its typical gravitational
radius
 $r_s$: $\delta M \sim (\Delta p )^2/M$ with $\Delta p \sim 1/r_s $.

Now we calculate the entropy of a rotating black hole
in five dimensions \cite{spn,vbd}. The angular momentum is characterized by
the eigenvalues on two orthogonal two-planes, $J_1,J_2$,
for example $J_1$ corresponds to rotations of the 12 plane and
$J_2$ to rotations of the 34 plane. 
In terms of the $J_3$ eigenvalues $J_L,J_R$ of the 
 SU(2)$_L\times$ SU(2)$_R \sim $SO(4) decomposition of the
spatial rotation group we find 
\be\la{angmom}
J_1 = J_L + J_R~~~~~~~~~~~~~~J_2 =J_L-J_R
\ee
As we mentioned above $J_R,J_L$ are carried by right and 
left movers respectively. They are also the eigenvalues of  U(1) 
appearing in the supersymmetry algebra.
States carrying U(1) eigenvalue $J$ have conformal 
weight bigger than $\Delta = {6 J^2/ c}$ where $c=6Q_1Q_5$ 
is the central 
charge. The states with minimum conformal weight 
correspond to states $ e^{i J \phi}|0\rangle$, 
where $j = { c\over 12 } \partial
\phi
$ is the U(1) current. In the total left moving
energy $N_L$ there is an amount $ J_L^2/Q_1Q_5$ which we are not 
free to distribute. It is fixed by the condition that
the system has angular momentum $J_L$, so the effective
number of left  movers that we are free to vary is $\tilde{N}_L =
N_L - J_L^2/Q_1Q_5 $. 
The same is true for the right movers, so that the entropy becomes
\be\la{rotentro}\ba{rl}
S = & 2\pi\sqrt{Q_1Q_5} (\sqrt{\tilde N_L} + \sqrt{\tilde N_R} ) =\\
=&
2\pi\sqrt{N_L Q_1Q_5 - J_L^2 }+
2\pi\sqrt{N_R Q_1Q_5 - J_R^2 }
\ea \ee
which agrees with the classical entropy formula of a rotating
black hole in the dilute gas regime \cite{vbd}.
Actually, in five dimensions we can have rotating BPS 
black holes by setting $N_R = J_R =0$, this implies $J_1=J_2$.
Again the corresponding formula agrees with the classical entropy
formula but the restriction to the dilute gas regime is no longer
necessary since the computation is protected by supersymmetry.

\section{Greybody factors}

The D-brane emission rate into massless scalars is given by
\refnew{ratefinal}. More precisely that is the emission into 
minimally coupled scalars, scalars that in the supergravity 
theory are not coupled to the vector fields that are excited in
the black hole background. 
According to the semiclassical analysis the emission rate should
be 
\be \la{hawrate}
d\Gamma = \sigma(\omega, r_0, \alpha, \gamma,\sigma ) 
{ 1  \over e^{ \omega\over T_H } -1 }
\ee
where $\sigma(\omega, r_0, \alpha, \gamma,\sigma ) $ is the absorption
cross section of the black hole which is a function of 
 the various parameters specifying the black hole solution 
\refnew{solnfd}. In the usual Schwarzschild black hole case the only 
scale in the solution would be the Schwarzschild radius $r_s$.
This emission rate \refnew{hawrate} has the same form for any body
emitting thermal radiation. The absorption cross section comes
in because of detailed balance: in order for the body
to be in equilibrium with a bath of radiation it has to absorb as much
as it
emits.
The prefactor in \refnew{hawrate} is usually called greybody factor,
since it is what makes bodies  grey instead of black.
At first sight, the semiclassical rate \refnew{hawrate} does not  seem to 
be in agreement with the D-brane rate \refnew{ratefinal} since one has
two exponential factors and the the other seems to have only one.
In order to see whether they really agree  we should calculate
the greybody factor. It turns out that the greybody factor is
precisely such that  these two calculations to agree. 
We now describe this calculation.

We consider the scattering of scalars from a five dimensional 
black hole  in the dilute gas limit $r_0,r_n \ll r_1, r_5$.
We also restrict to 
low energies satisfying $\omega \ll 1/r_1, 1/r_5 $ but there is no
restriction on ${\omega r_0 \over r_1 r_5 }$ or ${\omega r_n \over r_1
r_5}$, 
in other words, no restriction on $\omega/T_L,~\omega/T_R $.
 We follow the notation of \cite{asjm}, where further details of 
the geometry may also 
be found. The wave equation in this background becomes 
\be \la{waveeqn}
{h \over r^3} {d \over dr } r^3 h { d \phi \over dr }
 +
\omega^2 \lambda \phi =0,
\ee
where $\lambda,h $ are defined in \refnew{deff}.

We 
 divide space into a far region $r \gg r_1,r_5 $ and a near region
$r \ll  1/\omega$ and we will match the solutions in the overlapping
region.
In the far region, the equation is solved by
the Bessel functions
\be \la{solfar}
\phi = { 1 \over \rho } \left[ \alpha J_{\nu} (\rho) + \beta 
J_{-\nu }(\rho) \right].
\ee 
with $\rho =\omega r$ and $\nu^2 = 1 - \epsilon $, where $\epsilon =
\omega^2 (r_1^2 + r_5^2)$  is very small and we keep it to simplify
the
form of 
the intermediate equations but will disappear from the final answer.
   From the large $\rho$ behavior the incoming
flux is found to be
\be\la{flux}f_{in} = Im( \phi^* r^3 \partial_r \phi )
= { 1 \over 2  \pi\omega^2 } | \alpha e^{i \nu\pi/2} + 
\beta e^{-i \nu \pi/2} |^2.
\ee
On the other hand, the small $\rho$ behavior of the far region 
solution
is 
\be\la{smallrho} \ba{rl}
\phi = &{ 1 \over \rho } \left[
\alpha ({\rho\over 2})^{\nu}
( { 1 \over \Gamma(\nu+1 ) } - {\cal O}(\rho^2) ) \right.
\\ & \left. +
\beta ({\rho\over 2})^{-\nu} 
( { 1 \over \Gamma(-\nu +1) } - {\cal O}(\rho^2) ) \right].
\ea \ee
Now we turn to the solution in the near region $r \ll 1/\omega $. 
Defining $v = r_0^2/r^2$, the near region wave equation is
\be\la{nearone}
(1-v)^2 {d^2 \phi \over dv^2 } - (1-v) {d \phi \over dv }
+\left( C + {D \over v} + { E \over v^2 } \right) \phi =0
\ee
where 
\be\la{cdef}\ba{rl}
C = &\left( \omega r_n r_1 r_5 \over 2 r_0^2 \right)^2 ~,~~~~~~
D =  { \omega ^2  r_1^2 r_5^2 \over 4 r_0^2 }    
 + { \nu^2 -1 \over 4} ~, \\
E = & -{\nu^2 -1 \over 4}.
\ea \ee
Defining 
\be\la{defoff}
\phi = v^{- (\nu -1)/2} (1-v)^{-i {\omega \over 4 \pi T_H}} A  F 
\ee
with $A$ a constant, 
we find that the solution to \refnew{nearone} with only ingoing flux
at the horizon is given by \refnew{defoff} with the hypergeometric function
\be\la{hyperf}\ba{rl} 
F= & F(a,b,c; 1-v)
\\
a &= - \nu/2 +1/2 + i {\omega \over 4
\pi T_L}
\\
b &= - \nu/2 +1/2 + i {\omega \over 4
\pi T_R}
\\ 
c& =1 + i{\omega \over 2 \pi T_H}
\ea \ee 

 The behavior for small $v$ can be calculated by expressing the
hypergeometric
function \refnew{hyperf}, which depends on $1-v$, in terms
of hypergeometric functions depending on $v$ and then expanding in $v$.
Matching  this with
\refnew{smallrho}  
we find 
\be\la{defalpha}\ba{rl}
\alpha/2 =& A 
\left[ { \Gamma(1 + i{\omega \over 2 \pi T_H} ) \over
 \Gamma( 1 +i {\omega \over 4\pi T_L}  )
\Gamma( 1  + i {\omega \over 4\pi T_R} )} \right]~,
\\
\beta \ll & \alpha
\ea
\ee
The absorbed flux is 
\be\la{absflux}
f_{abs} = Im( \phi^* h r^3 \p_r \phi) 
= { \omega  r_0^2 \over 2 \pi T_H} |A|^2.
\ee
The absorption cross section for the radial problem is
given by the ratio  of the two fluxes \refnew{absflux} \refnew{flux}. The plane
wave cross section is obtained by multiplying by $4\pi/\omega^3$
\be \la{ crossradnew}
\sigma_{\rm abs} = 
{\pi^3 r_1^2r_5^2 \omega} { e^{ \omega \over T_H } -1
\over \left( e^{ \omega \over 2 T_L }  -1 \right)
\left( e^{ \omega \over 2 T_R }  -1 \right) }
\ee
where the exponential terms come from the gamma functions in
\refnew{defalpha}.
We see that it has precisely the right form to make the 
D-brane result \refnew{ratefinal} agree with the semiclassical
calculation 
\refnew{hawrate}.

These greybody factor calculations have been generalized to various 
cases. One possible generalization 
is to consider the emission of scalars that are not minimally coupled,
in some cases precise agreement is found \cite{fixedscalars} \cite{krt}.
From the D-brane point of view the difference between these scalars 
and the one that we have been considering is in the conformal weight
of the operator on the effective  SCFT that they couple to.
The minimally coupled scalars couple to operators of dimension (1,1) 
(like $\partial X \bar \partial X$) while the scalars in 
\cite{fixedscalars} couple to operators of conformal weights (2,2) or
(1,2) in \cite{krt}.There is still some puzzling disagreement for the case of
operators of weight (3,1) \cite{kk}, which hopefully will be resolved
soon!.
Another generalization is to consider the emission
of higher partial waves \cite{jmasrot,gubserrot,mathurrot}. 
These calculations of greybody factors shows that some of the features
of the near extremal geometry are encoded in  the dynamics of the 
1+1 dimensional gas (or the CFT). Since the wavelengths
of the particles we scatter are much bigger than the size of the 
black holes it is hard to get precise information about the 
metric. A more direct way to obtain information about the metric 
is by using D-brane probes \cite{dps}. In that approach one 
starts from D-branes in flat space and by integrating 
out the massive stretched open strings one obtains an action 
for the probe D-branes that is at some distance from the rest of the
branes. This action is then interpreted as the 
action of a D-brane in the presence of some classical supergravity 
background. This works to one loop \cite{dps,jmprobe} but the 
status of the higher loop contributions is unclear. 

The four dimensional black holes have a similar description
\cite{fourextremal,hlm,gubserfour}.

These results have clear implications in the information loss problem.

{\bf Acknowledgments}

I thank C. Callan, G. Horowitz, D. Lowe, A. Strominger, L. Susskind
and 
A. Peet, for many fruitful collaborations on this subject.

\end{document}